\documentclass[11pt]{article}
\usepackage{moriond,epsfig,latexsym}

\def\Journal#1#2#3#4{{#1} {\bf #2}, #3 (#4)}

\def\NPB{{\em Nucl. Phys.} B}
\def\PLB{{\em Phys. Lett.}  B}
\def\PRL{\em Phys. Rev. Lett.}
\def\PRD{{\em Phys. Rev.} D}

\def\be{\begin{equation}}
\def\ee{\end{equation}}
\newcommand{\LGB}{\mathcal{L}_\mathrm{GB}}
\newcommand{\mpl}{M_\mathrm{Pl}}
\newcommand{\mph}{M_{\phi}}
\newcommand{\dz}{\partial_z}
\begin{document}
\vspace*{4cm}
\title{
GRAVITY ON A DILATONIC GAUSS-BONNET BRANE WORLD}

\author{ STEPHEN C. DAVIS }

\address{
IPT, \'Ecole Polytechnique F\'ed\'erale de Lausanne,
CH--1015 Lausanne, Switzerland}

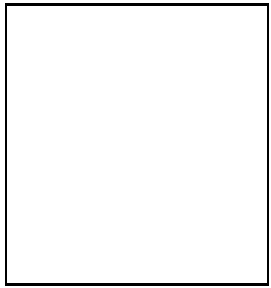
\maketitle\abstracts{
The effective four-dimensional, linearised gravity of a
Randall-Sundrum-like brane world model is analysed~\cite{GBgrav}. The
model includes higher order curvature terms (such as the
Gauss-Bonnet term) and a scalar field. The resulting brane worlds can
have better agreement with observations than the equivalent Einstein
gravity models.
}

\section{Brane Worlds and Higher Order Gravity}

In the second Randall-Sundrum (RS) brane world scenario~\cite{RSII}, we live on
3+1 dimensional brane embedded in a 4+1 dimensional bulk spacetime. As
a result of the warping of the fifth dimension, the effective
gravitational theory on the brane closely resembles that which is
observed in our universe (except at very small distances). In this
paper we will investigate an extended version of this scenario, which has
higher order gravity and a scalar field, $\phi$, in the bulk. We will
consider $Z_2$-symmetric solutions of the form $ds^2= e^{-2k|z|}
dx^2_4 +dz^2$ and $\phi=-\sigma |z|$ which is the simplest
generalisation of the RS model (in some sense $\sigma$ is
the scalar field equivalent of the warp factor $k$). To avoid bulk
singularities $\sigma$ needs to be positive.

In four dimensions, the gravitational field equations (for the vacuum)
are taken to be $G_{ab} + \Lambda g_{ab} = 0$. These can be derived by
looking for a rank 2 curvature tensor which (i) is symmetric, (ii) is
divergence free, and (iii) depends only on the metric and its first
two derivatives. In five
dimensions the above conditions are satisfied by
$G_{ab} + 2\alpha H_{ab} + \Lambda g_{ab} = 0$, where $H_{ab}$ is the
second order Lovelock tensor~\cite{Lovelock}. $H_{ab}$ can be obtained
from the variation of an action containing the Gauss-Bonnet term
\be
\LGB = R^{abcd} R_{abcd} - 4 R_{ab}R^{ab} + R^2 \ .
\ee
Energy momentum is conserved in the corresponding gravitational
theory and its vacuum is ghost-free (just as in Einstein gravity).
Note that $H_{ab}$ is the only quadratic curvature term
which satisfies the above three conditions. In four dimensions its
contribution to the field equations is trivial, and so it is usually
ignored.

The Gauss-Bonnet term also appears in low energy effective actions
derived from string theory. Since the brane worlds
are loosely motivated by string theory, it is particularly
natural for them to include higher gravity terms. String
theory also suggests that the bulk space will contain not only
gravity, but many other fields. One such field is the dilaton. For
simplicity we will include only include this one extra bulk scalar
field.

If a scalar field is present, it is natural to include higher order
scalar kinetic terms as well as the higher order curvature terms. We
will consider the  general second order contribution to the action (in
the string frame)
\be
\mathcal{L}_2 = 
c_1 \LGB - 16 c_2 G_{ab}\nabla^a\phi \nabla^b \phi 
+ 16 c_3  (\nabla\phi)^2 \nabla^2 \phi - 16 c_4 (\nabla\phi)^4  \ .
\label{L2}
\ee

For simplicity we will not consider higher than second order terms. In
this case the full bulk action is
\be
S_\mathrm{Bulk} = \frac{1}{2} \int d^5x  \sqrt{-g}  
e^{-2\phi}\left\{ R - 4\omega (\nabla \phi)^2
+ \mathcal{L}_2  - 2\Lambda \right\} \ .
\ee
The coefficients can be determined from origin of $\phi$. We will take
$\omega=-1$ and $c_i=\alpha$ which corresponds to the dilaton (with
some extra symmetries).

The brane can be treated as a boundary of the bulk spacetime. In this
case we need to add a Gibbons-Hawking boundary term (and
corresponding higher curvature terms) to the action
\be
S_\mathrm{brane} = - \int d^4x \sqrt{-h} e^{-2\phi} \, 
\left\{ 2K + \mathcal{L}_2^{(b)} + T \right\}
\ee
\be
\mathcal{L}_2^{(b)} = c_1 \LGB^{(b)} 
- 16 c_2 (K_{ab}- K h_{ab})D^a\phi D^b \phi
- 16 c_3 (n \! \cdot \! \nabla \phi) 
\left(\frac{1}{3}(n \! \cdot \! \nabla \phi)^2 + (D\phi)^2\right) 
\ee
where~\cite{Myers}
\be
\LGB^{(b)} 
= \frac{4}{3}( 3K K_{ac}K^{ac} - 2 K_{ac}K^{cb}K^a{}_b -K^3)
- 8 G^{(4)}_{ab}K^{ab} \ .
\ee

Variation of the action gives the generalised Israel junction 
conditions for the brane~\cite{BCGB}. These do not depend on the brane
thickness (this is not true for other second order gravity terms).

For the type of solutions we are considering, there are three
different solution branches for the bulk field equations
\be
\mathrm{(a)} \ \ k=0 \  \ , \ \
\mathrm{(b)} \ \ k = \sigma - \sqrt{\frac{1}{12\alpha} + \frac{\sigma^2}{3}}
\  \ , \ \
\mathrm{(c)} \ \ k = \sigma +\sqrt{\frac{1}{12\alpha}+\frac{\sigma^2}{3}} \ .
\label{sols}
\ee
The first is always valid, while the other two are only
possible when the higher order terms (\ref{L2}) are included in the
action.

\section{Linearised Brane World Gravity}

When analysing the effective four-dimensional brane gravity we need to
worry about perturbations of the brane position as well as bulk
metric. This can be addressed by also perturbing the coordinates, or
by using gauge in which brane stays at $z=0$~\cite{lingrav} (the
approach we will use).

Consider a general perturbation of the RS-like brane
world solutions with
\be
ds^2 = e^{-2k|z|}(\eta_{\mu \nu} + \gamma_{\mu \nu}) dx^\mu dx^\nu 
+ 2 v_\mu dx^\mu dz + (1+\psi) dz^2
\ee
and $\phi = -\sigma |z| + \varphi$, 
where $\gamma_{\mu \nu}$, $v_\mu$, $\psi$ and $\varphi$ are small.
It is useful to split $\gamma_{\mu\nu}$ into tensor and scalar parts
\be
\gamma_{\mu \nu} = 
\bar \gamma_{\mu \nu} + \frac{1}{4} \gamma \eta_{\mu \nu} 
+\frac{4}{3}c_\chi \left(\frac{1}{4}\eta_{\mu \nu}
  -\frac{\partial_\mu \partial_\nu}{\Box_4}\right)\chi \ ,
\label{gbar}
\ee
where $\gamma = \eta^{\mu \nu}\gamma_{\mu \nu}$ and
$\partial^\mu \bar \gamma_{\mu \nu} = 0$. The field $\chi$ will
be some linear combination of $\gamma$ and $\varphi$. If the brane
remains at $z=0$, the bulk field equations are satisfied by
$\psi= 2\dz (\chi -\varphi)/\sigma$ and 
$v_\mu = \partial_\mu (\chi -\varphi)/\sigma$.

\section{New Instabilities}

The graviton wave equation is obtained from the
remaining bulk field equations.
\be
\mu_\gamma \left(\dz^2 - 2(2k -\sigma)\dz 
+ f_\gamma^2 e^{2kz} \Box_4\right)\bar \gamma_{\mu \nu} = 0
\label{Bg}
\ee
where $\mu_\gamma = 1-4\alpha  [k^2 -4k\sigma+2\sigma^2]$ and
$f^2_\gamma = 1 - 8\alpha \sigma k/ \mu_\gamma$. If either of  $\mu_\gamma$ or
$f^2_\gamma$ is negative, the kinetic term in the effective bulk
action for $\bar \gamma_{\mu \nu}$ will have the wrong sign, and so
the theory will have ghosts~\cite{us}. This is not possible if
$\alpha=0$ (i.e.\ if the higher order gravity terms are absent).

The brane junction conditions imply
\be
\mu_\gamma \dz \bar \gamma_{\mu\nu}
+4 \alpha [k-2\sigma] \Box_4 \bar \gamma_{\mu\nu}
= -\left\{S_{\mu\nu} 
- \frac{1}{3}\left(\eta_{\mu\nu} 
- \frac{\partial_\mu \partial_\nu}{\Box_4}\right)S \right\}
\label{bg}
\ee
where $S_{\mu\nu}$ is the brane energy momentum tensor.

If $4\alpha[k-2\sigma] < 0$, then either the effective
four-dimensional Planck mass is negative, or the vacuum has
non-trivial solutions with spacelike momenta (i.e.\ tachyons, which
implies the solution is unstable). This occurs for the $k=0$ solution
branch~(\ref{sols}a) if $\alpha > 0$. The model would be stable if
$\alpha=0$ (but it does not give the correct gravitational laws). So
in this case the higher order corrections have destabilised the solution.
Scalar ghosts and tachyons are also possible in this type of model.

\section{New `Gauss-Bonnet' solutions}

When $\alpha > 0$ two new solution branches appear with
$k = \sigma \pm \sqrt{1/[12\alpha] + \sigma^2/3}$. The
graviton equations are eqs.~(\ref{Bg}) and (\ref{bg}), and
$\mu_\gamma = 8\alpha k (k-\sigma)$ and $f_\gamma^2 = (k-2\sigma)/(k-\sigma)$. 
Ghosts and tachyons are possible, and we find that the second solution
branch~(\ref{sols}b) is always unstable. For the rest of this article
we will only consider the third solution branch~(\ref{sols}c), which
is stable if $\sigma < 1/\sqrt{8\alpha}$.

Switching to Fourier space, the bulk graviton equation is solved by
\be
\bar \gamma_{\mu\nu} \propto e^{(2k-\sigma)z} 
K_{2-\sigma/k} \left(f_\gamma p e^{k z}/k\right)
\ee
for spacelike perturbations ($p=\sqrt{p^\mu p_\mu}$).

For $p \ll k/f_\gamma$ (which corresponds to larger distances)
\be
\dz \bar \gamma_{\mu\nu} \approx 
\frac{f_\gamma^2}{2(k-\sigma)}  \Box_4 \bar\gamma_{\mu\nu}
\label{gser}
\ee
so on the brane, the junction condition~(\ref{bg}) reduces to the
usual, four-dimensional, linearised Einstein equation at large
distances (just as in the RS model).  The extra $\Box_4
\bar\gamma_{\mu\nu}$ term in the junction condition gives
four-dimensional gravity at short distances too (unlike the RS
model). This significantly weakens the constraints on the model from
gravity experiments~\cite{nathalie}.

Unlike the RS model, we also have a scalar field to worry
about. The effective bulk scalar field is
\be
\chi = \frac{6\alpha k(k-\sigma)}{2\sigma +3k(1+4\alpha k^2)}
(8k \varphi - \sigma \gamma) \ .
\ee
The constant $c_\chi$ in eq.~(\ref{gbar}) is then
$-\sigma (3k-2\sigma)/[k(k-\sigma)]$.

The bulk equation for
$\chi$ is qualitatively similar to that of the graviton, but with
$\mu_\chi = 16\alpha k(3k-2\sigma)$ and $f_\chi^2 = 3k/(3k-2\sigma)$.

The remaining junction conditions are
\be
(3k-2\sigma) [12k^2(k-\sigma) + \alpha (3k-2\sigma)] \dz \chi
+12k^2\sigma \Box_4 \chi = \frac{S}{16}
\ee
\be
-4\alpha \sigma k (3k-2\sigma) \dz \chi
 +12\alpha k (k-\sigma) \Box_4 \chi
= \Box_4 \varphi \ .
\ee
These are all qualitatively similar to the graviton equations. Again we find
there are no scalar ghosts or tachyons for solution (c) if
$\sigma < 1/\sqrt{8\alpha}$. The scalar perturbations give
approximately four-dimensional brane gravity at large and small
distances, in the same way that the graviton perturbations do. However
all the coefficients in the above equations are different. We see that
the degeneracy between the behaviours of scalar and tensor modes has
broken. In particular we can have $f_\gamma \ll f_\chi$ if $\sigma$
is near $1/\sqrt{8\alpha}$, and so the two types of perturbation `feel'
the effects of the bulk differently. 

Using the solutions of the above field equations, and taking
appropriate series and asymptotic expansions, we obtain (to
leading order) linearised Brans-Dicke gravity on brane.
\be
\mathcal{G}_{\mu\nu} 
- 2(\eta_{\mu\nu} \Box_4 - \partial_\mu \partial_\nu) \tilde \varphi
\approx  \mpl^{-2}S_{\mu\nu} \ , \ \
-2\Box_4 \tilde \varphi \approx \mph^{-2} S \ ,
\ee
where $\mathcal{G}_{\mu\nu}$ is the linearised Einstein tensor
corresponding to $\gamma_{\mu\nu}$ and $\tilde
\varphi(\gamma,\varphi)$ is the effective four-dimensional scalar. The
effective (distance dependant) coupling strengths of gravity and the
scalar are respectively $\mpl^{-2}$ and $\mph^{-2}$.

If $p \ll k/f_\chi$ (large distances), we find
\be
\mpl^2 = 8\alpha f_\gamma^2 (2k-\sigma) \ , \ \
\mph^2 = 8\alpha (3k-2\sigma) \ ,
\ee
so we can have $\mph \gg \mpl$ if the solution is fine-tuned to have 
$f_\gamma \ll 1$. Hence we can potentially avoid
conflict with solar system constraints~\cite{Damour}. Note that this
is possible despite the fact that the coupling strengths of gravity
and $\phi$ are the same in the underlying five-dimensional theory. 

At medium ($k/f_\gamma \gg p \gg k/f_\chi$)  and short ($p \gg k/f_\gamma$)
distance scales, we find $\mph^2 \leq 3\mpl^2$. But this is not a
problem, since short distance constraints on Brans-Dicke gravity are weak.

It is interesting to note that the series expansion~(\ref{gser}) used
to determine the effective large distance gravity is not valid if
$\sigma > k$. If $2k > \sigma > k >0$ we find $\dz \bar
\gamma_{\mu\nu} \sim p^{4-2\sigma/k} \bar \gamma_{\mu\nu}$
instead. The corresponding large distance Newton potential will then
have a non-standard $r^{1-2\sigma/k}$ behaviour. This is possible
for the second solution branch~(\ref{sols}b) when
$\alpha<0$. Unfortunately it is has bulk ghosts and so is unstable. It
may be possible that by using a different choice of coefficients in
eq.~(\ref{L2}), a stable version of this type of solution can be found.

\section*{Acknowledgements}
I wish to thank the Swiss Science Foundation for financial support,
and the conference organisers for giving me the opportunity to speak at
such an enjoyable conference.


\section*{References}
\begin{thebibliography}{99}
\bibitem{GBgrav}
S.~C.~Davis, hep-th/0402151.
\bibitem{RSII}
L.~Randall and R.~Sundrum,
\Journal{\PRL}{83}{4690}{1999}  [hep-th/9906064].

\bibitem{Lovelock}
D.~Lovelock,
\Journal{{\em J. Math. Phys.}}{12}{498}{1971}. 

\bibitem{Myers}
R.~C.~Myers,
\Journal{\PRD}{36}{392}{1987}.

\bibitem{BCGB}
S.~C.~Davis,
\Journal{\PRD}{67}{024030}{2003} [hep-th/0208205]. \\
E.~Gravanis and S.~Willison,
\Journal{\PLB}{562}{118}{2003} [hep-th/0209076].

\bibitem{lingrav}
I.~Y.~Aref'eva, M.~G.~Ivanov, W.~Muck, K.~S.~Viswanathan and I.~V.~Volovich,
\Journal{\NPB}{590}{273}{2000} [hep-th/0004114]; \\
H.~Collins and B.~Holdom,
\Journal{\PRD}{62}{124008}{2000}  [hep-th/0006158].

\bibitem{us}
C.~Charmousis, S.~C.~Davis and J.~F.~Dufaux,
\Journal{\em{JHEP}}{0312}{029}{2003}  [hep-th/0309083].

\bibitem{nathalie}
N.~Deruelle and M.~Sasaki,
\Journal{{\em Prog. Theor. Phys.}}{110}{441}{2003} [gr-qc/0306032].

\bibitem{Damour}
T.~Damour and K.~Nordtvedt,
\Journal{\PRD}{48}{3436}{1993}, (and references therein). 


\end{thebibliography}
\end{document}